\documentclass[12pt,aps,amsfonts,nofootinbib,superscriptaddress]{revtex4-1}
\usepackage{graphicx}
\usepackage{amsmath}
\usepackage{ulem}
\usepackage{color}
\begin{document}
\newcommand{\bb}{\begin{equation}}
\newcommand{\ee}{\end{equation}}
\newcommand{\eqb}{\begin{eqnarray}}
\newcommand{\eqf}{\end{eqnarray}}
\newcommand{\1}{\'{\i}}
\newcommand{\journal}[4]{{\it #1,\/} {\bf #2}, #3 (#4)}

\title{Photoproduction total cross section and shower development}

\author{F. Cornet}
\affiliation{Departamento de F\'{\i}sica Te\'orica y del Cosmos and
Centro Andaluz de F\'{\i}sica de Part\'{\i}culas, Universidad de
Granada, E-18071 Granada, Spain.}

\author{C. A. Garc\'{\i}a Canal}
\affiliation{Departamento de F\'{\i}sica, Universidad Nacional de La
Plata, IFLP, CONICET, C. C. 67, 1900 La Plata, Argentina.}

\author{A. Grau}
\affiliation{Departamento de F\'{\i}sica Te\'orica y del Cosmos and
Centro Andaluz de F\'{\i}sica de Part\'{\i}culas, Universidad de
Granada, E-18071 Granada, Spain.}

\author{G. Pancheri}
\affiliation{INFN Frascati National Laboratories, Via Enrico Fermi
40, I-00044 Frascati, Italy.}

\author{S. J. Sciutto}
\affiliation{Departamento de F\'{\i}sica, Universidad Nacional de La
Plata, IFLP, CONICET, C. C. 67, 1900 La Plata, Argentina.}

\begin{abstract}
 The total photoproduction cross section at ultra-high
energies is obtained using a model based on 
QCD minijets and soft-gluon resummation and the ansatz
that infrared gluons limit the rise of total cross sections.
 This cross section is introduced into the Monte
Carlo system AIRES to simulate extended air-showers initiated by
cosmic ray photons. The impact of the new photoproduction
cross section on common shower observables, especially those related
to muon production, is compared with previous results.
\end{abstract}

\maketitle

\section{Introduction}

The determination of the composition of ultra-high energy (UHE),
i.e. with energies lager than $10^{18} \; eV$, cosmic rays is an
important open problem in cosmic ray physics. A good knowledge of the
percentage of protons, heavy nuclei and photons hitting the atmosphere
can provide important clues to understand the origin of those cosmic
rays and their acceleration mechanism. The Pierre Auger Observatory
\cite{PAO} has devoted big efforts to this end and they have recently
determined that the composition of UHE cosmic rays varies from mainly
protons at $E = 10^{17.5} \; eV$ to have an important presence of
heavy nuclei at $E =10^{19.5} \; eV$ \cite{Aab:2014aea}. However, as
no photons have been found up to now, the following bounds on the
fraction of photons arriving to Earth have been set: $3.8 \%$, $2.4
\%$, $3.5 \%$ and $11.7 \%$ for photon energies above $2 \times
10^{18} \; eV$, $3 \times 10^{18} \; eV$, $5 \times 10^{18} \; eV$ and
$10 \times 10^{18} \; eV$, respectively \cite{Abraham:2009qb}. An
important parameter to obtain these bounds is the photon-proton total
cross section from which one can estimate the photon-nucleus total
cross section.

There are no experimental values for $\sigma_{total}^{\gamma p}$ at the very high 
cosmic ray  energies, so one has to rely on extrapolations of accelerator
data that in the case of  photoproduction 
are limited to $\sqrt{s}\lesssim 200\ GeV $. Consequently the extrapolation 
to higher energies leaves considerable uncertainties. A possibility 
is to use recent LHC data, right now up to  $\sqrt{s}=8\ TeV$, and use models 
describing both photoproduction and $proton-proton$ scattering  to infer,  
from $pp$ data,  the higher energy behavior of $\sigma^{\gamma p }_{total}$. 
Notice that  the release of LHC data on the total proton-proton 
cross section, has led most current  hadronic  models  to slight 
adjustments of the model parameters.  The reason follows from the fact 
that lower energy data on hadron-hadron scattering, notably at 
$\sqrt{s}=540\ GeV$ and $1800\  GeV$ had an uncertainty of $10\%$ or more. 
Thus, often, a band, rather than a single curve, was provided. After
TOTEM data appeared  \cite{TOTEM7,TOTEM8}, the models could be sharpened 
taking into account  the much smaller error reported. On its turn, 
this sharpened tuning could be used for the high energy extrapolation 
of $\sigma^{\gamma p}_{total}$.

Here we shall follow the mentioned procedure. The updating of previous
predictions \cite{ourphoton} 
for $\sigma^{\gamma p}_{total}$ on the basis of LHC data, are 
then used as input to the
AIRES \cite{AIRES} system to simulate extended air showers initiated
by photons.

 The model for the total cross section,which we apply here, includes basic
 QCD inputs such as the parton densities obtained from experiments and
 well known QCD subprocess cross sections 
 \cite{ourproton}. A few non-perturbative
 parameters are also included. These ingredients allowed the search
 for the effects of the hadronic structure of the photon through the
 analysis of the total cross sections in which they are involved.

 In summary, the model developed in \cite{ourproton} 
is based upon the use of:
 \begin{itemize}
 \item QCD mini-jets to drive the rise of the total cross section in
 the asymptotic regime;
 \item The eikonal representation for the total cross section using a purely
 imaginary {\it overlap function}, obtained from mini-jet QCD cross sections;
 \item The impact parameter distribution, input for the eikonal,
 obtained from the Fourier transform of the re-summed soft gluon
 transverse momentum distribution;
 \item The resummation of soft gluon emission  down to zero
 momentum.
 \end{itemize}

The last element is the specific feature of this model, hereafter
called BN-model from the well known Bloch and Nordsieck \cite{BN} study
of the infrared catastrophe, which occurs in electrodynamics when the
soft photon momentum goes to zero. In the model, the resummation of
QCD soft gluons is applied, and covers the region where
$k_t^{gluon}\rightarrow 0$ with an ansatz, discussed below.  One
should notice that the main difference of this proposal with respect
to other mini-jet models comes from the energy dependence of the impact
parameter distribution and from soft gluon $k_t$-resummation extended
to zero momentum modes. The model to be presented in the next section
probes into this region.

\section{The Bloch-Nordsieck (BN) model for total cross sections}

  In this section we shall update previous results from the BN model
  for the 
$\gamma-p$ total cross section \cite{ourphoton}, which will be
  input to the AIRES simulation program.

\begin{figure}
\begin{center}
\includegraphics[width=14cm]{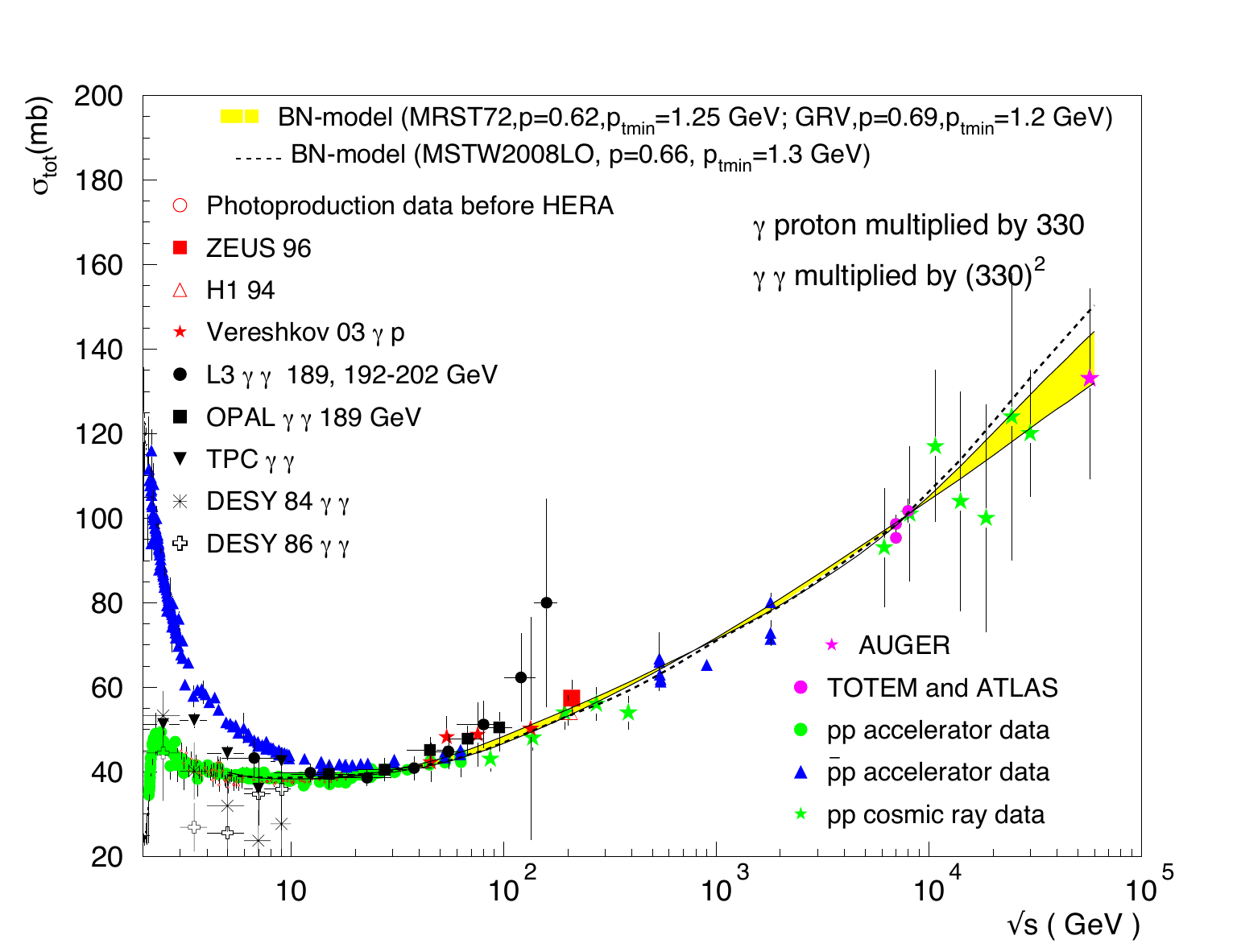}
\end{center}
\caption{ Total proton-proton, $p{\bar p}, \gamma p, \gamma \gamma$  
cross sections, 
normalized { at low energy} so as to show common features. This figure is updated from \cite{ourphoton}  to include recent data. The yellow band 
and the dashed line are obtained from  a description of $pp$ scattering with 
  an eikonal  model inclusive  of  mini-jets and  soft gluon resummation,
  described in the text and called  BN model.  }
  \label{fig:all}
\end{figure}

The BN  model \cite{ourproton} is based on two features of all
hadronic cross sections:
\begin{itemize}
\item As the c.m. energy increases from fixed target experiments to
  those at colliders, both purely hadronic and with photons, such as
  HERA ($\gamma p$), all total
  cross sections first decrease and then, around $\sqrt{s}=10-20
  \ GeV$ for the $pp$ case, start increasing. Mini-jet models
  attribute this transition to the onset of hard and semi-hard
  parton-parton collisions, which can be described by perturbative
  QCD. Such a suggestion was advanced long time ago \cite{cline} when
  proton-proton scattering at the CERN ISR \cite{ISR} confirmed the
  rise which cosmic ray experiments had already seen \cite{CRCold}.
  The large errors affecting the cosmic ray experimental data had cast
  uncertainty on a definite conclusion, but the ISR measurements
  definitively confirmed the rise, which was soon interpreted as a
  clear indication of the composite parton picture we are familiar
  with today. The role of mini-jets both in minimum bias physics \cite{Gaisser:1988ra} and in the rise of the total cross section \cite{Durand:1998ax} was then further developed and is input to many simulation programs. 
\item The observed rise, which may initially be considered to follow a
  power law, must obey the limitations of the Froissart bound, namely
\begin{equation}
\sigma_{total}(s) \simeq [\ln s]^2 \ \ \ \ 
 as \ \ \ s \uparrow \ \ \ \ at \ \ \ most
\end{equation}
where the bound is connected to the existence of a 
cut-off in impact parameter space \cite{froissart}.
\end{itemize}
 Both features are embedded into figure \ref{fig:all}, where 
the proton
 and photon cross sections are shown together, normalized at low
 energy, to highlight their common features. The yellow band,
 superimposed to the data, comes from the BN model we shall describe
 below.

  The first feature of the total cross section, i.e. the rise at high
  energy, is obtained in mini-jet models, through a perturbative
  calculation based on the QCD jet cross section, namely
\begin{equation}
 \sigma^{AB}_{\rm jet} (s,p_{tmin})=
\int_{p_{tmin}}^{\sqrt{s}/2} d p_t \int_{4
p_t^2/s}^1 d x_1  \int_{4 p_t^2/(x_1 s)}^1 d x_2 \times
\sum_{i,j,k,l}
f_{i|A}(x_1,p_t^2) f_{j|B}(x_2,p_t^2)
  \frac { d \hat{\sigma}_{ij}^{ kl}(\hat{s})} {d p_t},
  \label{minijets}
  \end{equation}
with $A,B = p, \bar p,\gamma$, and   $d \hat{\sigma}_{ij}^{ kl}(\hat{s}) /d p_t$  is the parton-parton differential cross section, { calculable from QCD, with  the running coupling constant of asymptotic freedom expression}. The parameter
$p_{tmin}
\approx 1-2$ GeV separates hard processes, for which one can use a
perturbative QCD description,  from the soft ones which dominate at low
c.m. energy of the scattering hadrons. The mini-jet cross section gets its name because is
dominated by low-$p_t$ processes, which cannot be identified by jet
finding algorithms, but can still be perturbatively calculated using
parton-parton sub-processes and DGLAP evoluted LO Partonic 
Density
Functions (PDFs) $f_{i|A}$, such as GRV \cite{GRV}, MRST \cite{MRST}, CTEQ
\cite{CTEQ} for the proton or GRV\cite{GRVPHO}, GRS \cite{GRS} and
CJKL \cite{CJKL} for the photon. 

 The expression of
equation (\ref{minijets}) gives a mini-jet cross section which rises very
fast with energy.
In order to ensure unitarity \cite{Durand:1998ax}, the mini jet cross section  is embedded in an
eikonal representation, whose implementation requires modeling the
impact parameter space of the colliding hadrons. We notice that, in order to obey   the limitations imposed by the Froissart bound \cite{froissart}, such modeling 
should  include   a large distance cut-off. 
In the BN model this is obtained by means of soft gluon resummation down to zero momentum gluons.

 In eikonal mini-jet models   one starts with
\begin{equation}
 \sigma _{tot}  = 2\int {d^2 {\bf b}} \left[ {1 - e^{ - n(b,s)/2} } \right]
 \label{sigtot}
\end{equation}
where $b$ is the impact parameter and the real part of the eikonal has been neglected. This is a good approximation at high energy.
The average number of collisions $n(b,s)$ can be split into  a soft contribution which will
be parameterized with a suitable non-perturbative expression, and a
perturbative (pQCD) term where both hard and soft gluon 
{emission
contribute, namely $n(b,s)= n_{soft}
(b,s) + n_{hard} (b,s)$.
In the mini-jet model of \cite{ourproton} the authors propose 
 \begin{equation}
 n_{hard} (b,s) = A_{BN}(b,s)\sigma _{jet} (s) \\
 \label{nhard}
 \end{equation}
with
\begin{equation}
A_{BN}(b,s)=N \int d^2{\bf  K_{\perp}}\  e^{-i{\bf K_\perp\cdot b}}
 {{d^2P({\bf K_\perp})}\over{d^2 {\bf K_\perp}}}={{e^{-h( b,q_{max})}}\over
 {\int d^2{\bf b} \ e^{-h( b,q_{max})}
 }}\label{eq:abn}
\end{equation}
and
\begin{equation}
h( b,q_{max}) =  \frac{16}{3}\int_0^{q_{max} }
 {{ \alpha_s(k_t^2) }\over{\pi}}{{d
 k_t}\over{k_t}}\log{{2q_{\max}}\over{k_t}}[1-J_0(k_tb)]
 \label{hdb}
\end{equation}
 The physical content of equation (\ref{eq:abn}) is as follows:
$A_{BN}(b,s)$ is obtained as the Fourier transform of the probability $d^2P({\bf K}_\perp)/d^2 {\bf K}_\perp$ that, in a collision between two collinear partons, soft gluon emission gives rise to an overall   transverse momentum unbalance ${\bf K}_\perp$. This probability  can be  calculated through    resummation of all soft gluons emitted
in an otherwise collinear  parton-parton collision. This  leads to the
exponentiation of the integrated single soft gluon spectrum, given by
the function $h( b,q_{max})$ of equation (\ref{hdb}). 

The  energy parameter $q_{max}$ represents the maximum
transverse momentum for {\it single} gluon emission and embeds the 
kinematics of  the process
\begin{equation}
parton_1 + parton_2\rightarrow jet_1+jet_2 +initial\  state \ emitted \ gluon
\end{equation}
Its calculation is detailed in \cite{corsetti1996} and  follows  the original formulation of \cite{greco}. As shown in  \cite{corsetti1996}, the energy parameter $q_{max}$ depends linearly on the
$p_t$ of the final state partons which, at leading order, are
described as hadronizing in two jets. This is  a
 description, which should hold at LO and upon averaging over
all densities and sub-processes. Thus $q_{max}$ depends on the  PDFs chosen  
for the
calculation of the mini jet cross section.   A description of 
$q_{max}$ and its dependence upon energy, PDFs and $p_{tmin}$ was recently presented in \cite{ourlastPRD}  for purely proton
processes for the  photon case, it can be found  in \cite{ourphoton}.  
 
  Because the acollinearity introduced by soft gluon emission reduces
  the cross section, the distribution of equation (\ref{eq:abn}) can
give a cut-off in ${\bf b}$-space, dynamically generated by soft gluon
emission and thus can reduce the very fast rise due to the mini-jet
cross section.  This effect is energy dependent, and increasing
through the energy parameter $q_{max}$, with the strength of the
cut-off depending on the infrared region. This region is crucial
to the calculation of the very large impact parameter processes,
dominating all total cross sections.

In the BN model, the zero momentum gluon contribution 
is implemented
by means of a singular but integrable behavior of the quark-gluon
coupling constant in the infrared region, characterized by a
singularity parameter $1/2 < p < 1$. 
 For details we refer the reader to references \cite{ourfroissart,ourproton}. To fit this parameter, in this model one uses the expression
 \begin{equation}
 \alpha_s(k_t^2)=\frac{12\pi}{33-2N_f}
  \frac{p}{
 \ln [
 1+p
 (
 \frac{k_t}
 {\Lambda_{QCD}}
 )^{2p}
 ]
 } \xrightarrow{k_t\to 0} \frac{12\pi}{33-2N_f}(\frac{\Lambda_{QCD}}{k_t})^{2p} .\label{eq:alphas}
 \end{equation}
 This expression reduces to the usual asymptotic freedom expression for $k_t>> \Lambda_{QCD}$, while 
 the  singular behavior for $k_t\rightarrow 0$ leads to a cut-off in impact parameter space, which is exponential for $p=1/2$, almost  gaussian for $p\lesssim 1$, and provides a mechanism for the implementation of the Froissart bound. As 
it was
shown in \cite{ourfroissart}, the singular  behavior of the coupling constant in the infrared limit leads to a large  impact parameter behavior such as  $A_{BN}(b,s)\simeq exp[-(b{\bar \Lambda})^{2p}]$. When coupled with the strong rise of  the mini jet cross section $\sigma_{jet}\simeq s^{\epsilon}$,  with $\epsilon \simeq 0.3-0.4$, one obtains
 \begin{equation}
 \sigma^{pp}_{total} \rightarrow [\ln s]^{1/p}  \ \ \ \ \ \sqrt{s}\rightarrow \infty
\end{equation}
The singularity parameter $p$ together with  $p_{tmin}$ and the choice of PDFs { determine completely $n_{hard} (b,s)$} and constitute the {\it high energy parameter set} of the BN model.
 The  choice of LO rather than next-to-leading order densities is discussed in \cite{ourproton} and follows from the  ansatz that resummation takes into account most of the major next-to-leading order contributions. The remaining uncertainty, from non-resummed  finite radiative corrections,  is included in the parameters $p$ and $p_{tmin}$, which are determined phenomenologically. 

Prior to the start of LHC, with the above inputs, and a phenomenological parameterization of the
low energy region, the BN model gave a good description of $pp$ and
$p{\bar p}$ total cross sections, predicting
$\sigma(\sqrt{s}=14\ TeV)= 100 \pm 12 \ mb$ \cite{ourlastPLB}. The 
  error corresponds to different  choices of the PDFs and of the
 parameter set $\{p,p_{tmin}\}$ and reflects  the difficulty of determining the optimal sets of parameters, because of the large error from $p {\bar p}$ measurements at $Sp {\bar p}S$ and TeVatron energies.

The measurement of the total
$pp$ cross section at LHC at energies $\sqrt{s}=7$ and $8\ TeV$ has 
allowed to reduce the  parametric uncertainties present in most models.
In figure \ref{fig:all} we present our updated analysis, with a band 
corresponding to the predictions for $pp$ obtained with  two different sets  
of LO PDFs, MRST and GRV.
 The yellow band   shows how the BN  model
accommodates recent results for $\sigma^{pp}_{total}$, including the
extraction of the $pp$ cross sections from cosmic ray measurement by
the AUGER collaboration, at $\sqrt{s}=57\ TeV$ \cite{AUGER2012}. With the choice of
parameters as indicated in the figure and the set of MRST densities from 
\cite{MRST}, the value expected at $\sqrt{s}=14\ TeV$ is
$\sigma_{total}^{pp}=112.24\ mb$. In addition, we also plot $pp$ results 
obtained in \cite{ourlastPRD} using more recent PDFs, such as  MSTW08, 
indicated by the dashed line.
Using older  LO densities,
such as GRV \cite{GRV}, one can also obtain a good description i.e.
$\sigma_{total}^{pp}=109.3\ mb$, namely, once the TOTEM, ATLAS 
and AUGER points (with their errors) are included in the description, the 
results, for different densities, are rather stable up to LHC energies. 
However, it must be pointed out that  beyond  the  
present LHC  range ($\sqrt{s}=7,\ 8\ TeV$), there is
  a band 
of uncertainty in the model predictions,  which corresponds to different 
extrapolations of the low $x$ behavior of the densities as the energy 
increases.  
 When the next LHC data will be available, this band will hopefully be narrowed further. In the meanwhile, in the
update of our $\gamma p$ results to be described next, we shall use for 
the proton 
the same two sets of LO PDFs, MRST and GRV,  used by  our code in \cite{ourphoton}. This choice may be modified in future applications.

 We  can now update the model for $\gamma p$, which had been proposed before the LHC. 
  In   \cite{ourphoton}, following the model proposed by Fletcher, Gaisser and Halzen in \cite{Fletcher},  the BN 
model  had been  applied to photoproduction   
with the following minimal
modifications :
\begin{eqnarray}
 \sigma _{tot}^{\gamma p}  = 2P_{had}\int {d^2 b}
 \left[ {1 - e^{ - n^{\gamma p}(b,s)/2} } \right]
\label{eq:gamp}\\
P_{had}=\sum_{V=\rho,\omega,\phi}{{4\pi \alpha}\over{f_V^2}}
\label{eq:phad}\\
n^{\gamma p}(b,s)=n_{soft}^{\gamma p}(b,s) +n^{\gamma
  p}_{hard}(b,s)\nonumber \\
 = n_{soft}^{\gamma p}(b,s) + A(b,s) \sigma_{jet}^{\gamma p}(s)/P_{had}\\
n_{soft}^{\gamma p}(b,s)= {{2}\over{3}}  n_{soft}^{p p}(b,s)
\end{eqnarray}
The extension to photon process requires the probability $P_{had}$
that the photon behaves like a hadron \cite{collins,Fletcher}. This
quantity is non perturbative and could have some mild energy
dependence. However, to minimize the parameters, it was taken
to be a constant, estimating it through
Vector Meson Dominance. In the analysis of \cite{ourphoton}, the value
$P_{had}=\frac{1}{240}$ was used.

To determine the $\gamma p$ cross section that will be used as input
to the AIRES shower simulation program \cite{AIRES}, we use
equation (\ref{eq:gamp}). We update 
the values of the model parameters to take
into account the impact of the recent LHC \cite{TOTEM7,TOTEM8,ATLAS}
and AUGER Observatory \cite{AUGER2012} results on the $pp$
cross section, which have appeared after the original analysis of
\cite{ourphoton}. 

 The result  is shown in figure \ref{fig:gamp}, with a band of values
 for $\sigma_{total}^{\gamma p}$ and  it is compared  
 with fits by Block and collaborators  \cite{block2004,block2014},
which impose a
Froissart-limit saturating the
high energy behavior. The band reflects results from the BN model with two different PDF
 sets. 
\begin{figure}
\begin{center}
\includegraphics[width=14cm]{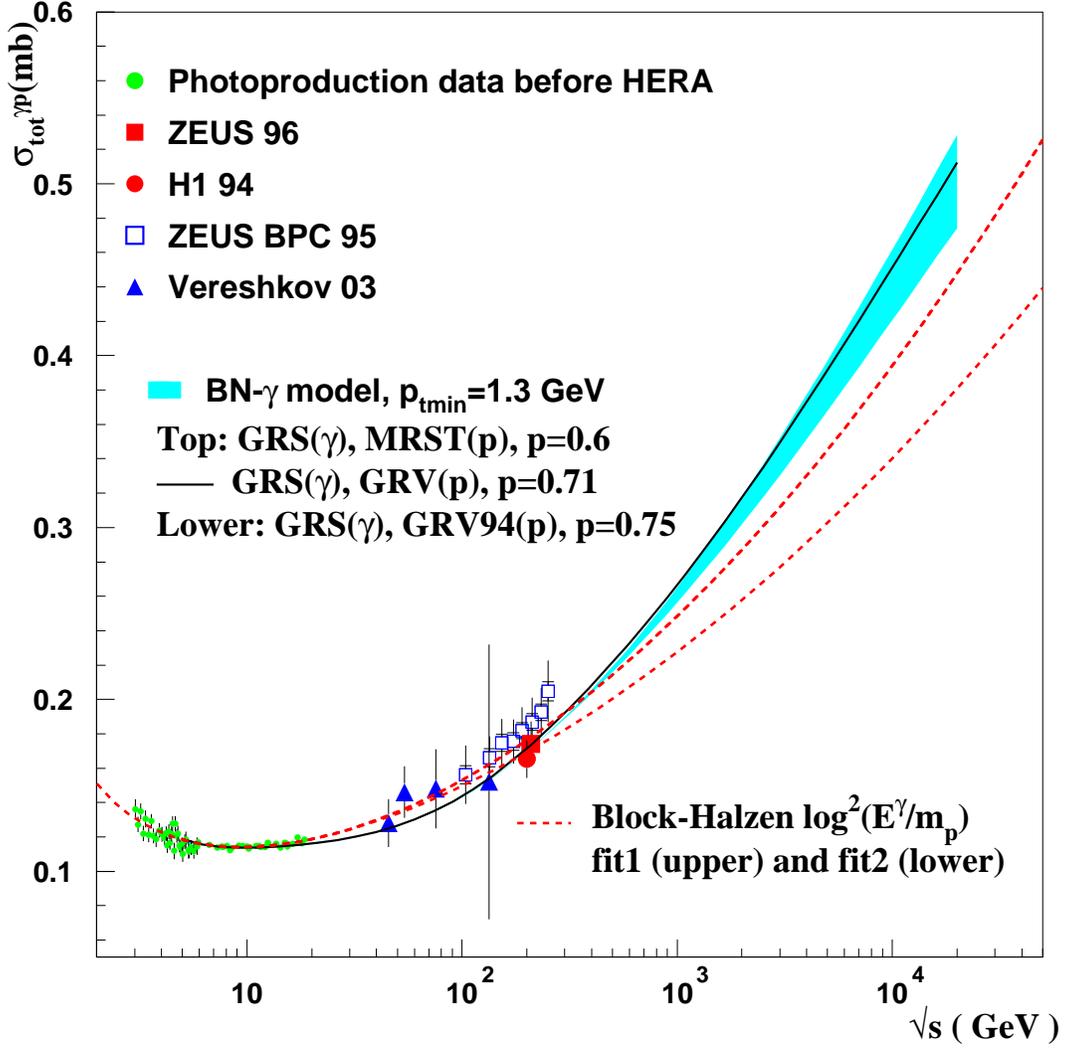}
\end{center}
\caption{ Total photoproduction cross section and its description with
  the BN model, and with the analytic model of ref. \cite{block2004,block2014}. }
\label{fig:gamp}
\end{figure}
For this application of the BN model, labeled as $BN-\gamma$, we have
used GRS densities for the photon, and the two PDF sets for the proton
as in figure \ref{fig:all}. The most recent type of PDFs, MSTW2008
\cite{MSTW2008}, gives results similar to MRST for $pp$, and have
similar uncertainties in the extrapolation to higher, AUGER type,
energies \cite{ourlastPRD}. Other parameters are obtained by comparison with the
 proton-proton results, within a few percent from those which give the
 yellow band for proton-proton in figure \ref{fig:all}.

\begin{figure}
\centering
\includegraphics{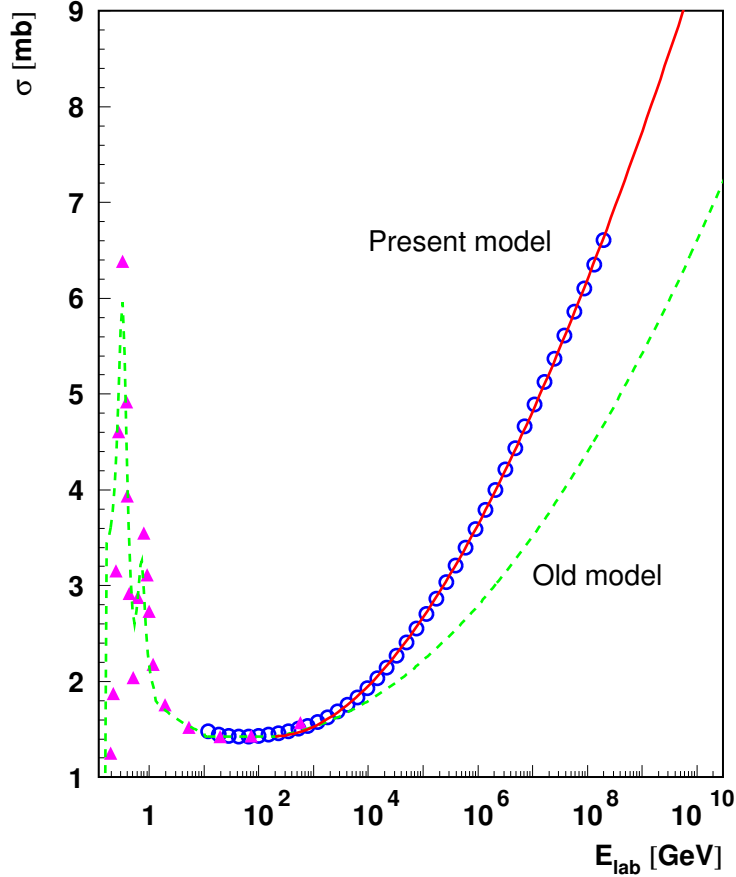}
\caption{%
  Photon-air nucleus cross sections used in our simulations,
  plotted versus the
photon lab energy. The triangles correspond to
  experimental data taken from reference \cite{PDG}. The open circles
  correspond to the present model, and the solid
  line corresponds to a fit to these data, valid for energies greater
  than 200 GeV. The dashed line corresponds to the cross sections used
  in AIRES.}
\label{fig:gammacrsn}
\end{figure}

\section{Simulation Results}

We have performed simulations of extended air showers using the AIRES
system \cite{AIRES} linked to the package QGSJET-II \cite{QGSJET}
for processing high energy hadronic interactions. We have run two sets
of simulations, namely, (1) using the cross sections for photonuclear
reactions at energies greater than 200 GeV that are provided with the
currently public version of AIRES; and (2) replacing those cross
sections by the ones corresponding to the present model. More
precisely, we have chosen to use the
  $\gamma p$ cross sections corresponding to the upper curve of the
blue band in figure \ref{fig:gamp} and fit 2 of \cite{block2004,block2014},
appropriately scaled to give the
  photon-air cross section required in AIRES. We are going to refer
to sets (1) and (2) as ``old model'' and ``present model'',
respectively.  In figure \ref{fig:gammacrsn} the gamma-air nucleus
cross sections corresponding to both sets are displayed as a function of the
photon lab energy. The triangles correspond to experimental data
taken from reference \cite{PDG}, while the open circles correspond to
numerical calculations using the present model, and is valid for
energies greater than 200 GeV. The dashed line corresponds to the up
to now standard cross sections implemented in AIRES, the ``old model'' 
\cite{block2004}. Notice that for energies below 200 GeV we always use
the same cross sections, which are calculated from fits to
experimental data.

\begin{figure}
\centering
\includegraphics[width=2.8in]{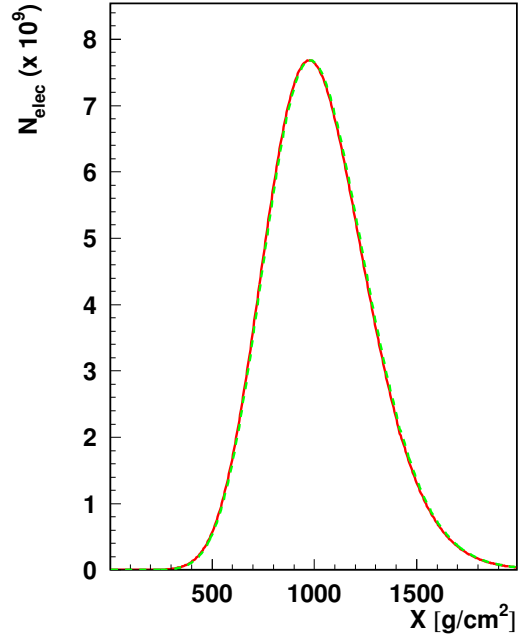}
\caption{Longitudinal development of electrons and positrons for
  $10^{19}$ eV photon showers inclined 60 degrees; ground at sea
  level. The solid (dashed) line corresponds to simulations with the
  present (old) model for the photonuclear cross section.}
\label{fig:longiepm}
\end{figure}

\begin{figure}
\begin{center}
\begin{tabular}{cc}
\includegraphics[width=2.8in]{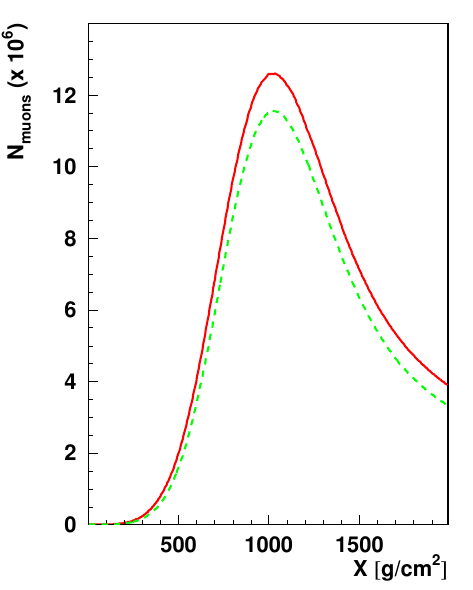} &
\includegraphics[width=2.8in]{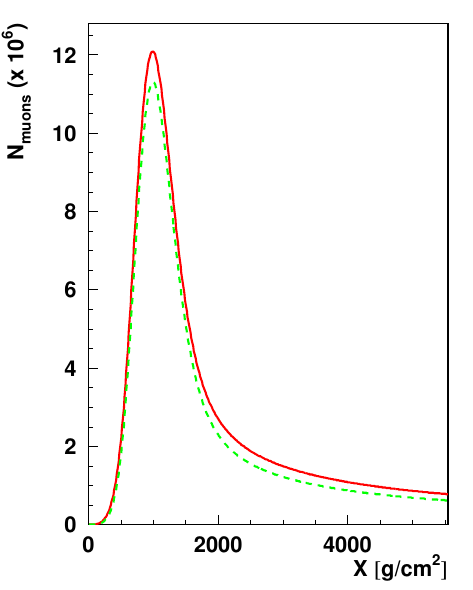}
\end{tabular}
\end{center}
\caption{Longitudinal development of muons for $10^{19}$ eV photon
  showers inclined 60 (left) and 80 (right) degrees.
 The solid (dashed) line corresponds to simulations with the
  present (old) model for photonuclear cross section.}
\label{fig:longimu}
\end{figure}

An important case to study the impact of changing the photonuclear
cross sections at high energy is the case of showers initiated by
photons. In such showers, the photonuclear reactions constitute the
main channel for hadron  production, 
which in turn are responsible
for the production of muons, mainly via pion decay. It is a well known
fact that showers initiated by photons have noticeably less muons
than showers initiated by hadrons, and this is one of the features
used to discriminate
photonic from hadronic showers.

For reasons of brevity, in this paper we present results only for the
very representative case of $10^{19}$ eV gamma showers. At this
primary energy, geomagnetic conversion \cite{Billoir2001} is not
frequent, thus allowing photons to enter the atmosphere unconverted,
and initiate normally the shower development. We have taken in most of
our simulations a ground altitude of 1400 meters above sea level,
corresponding to the altitude of current Cosmic Ray
Observatories. In a few cases we have used for convenience a sea level
ground altitude; this is explicitly indicated when it corresponds.

\begin{figure}
\centering
\begin{tabular}{cc}
\includegraphics[width=2.8in]{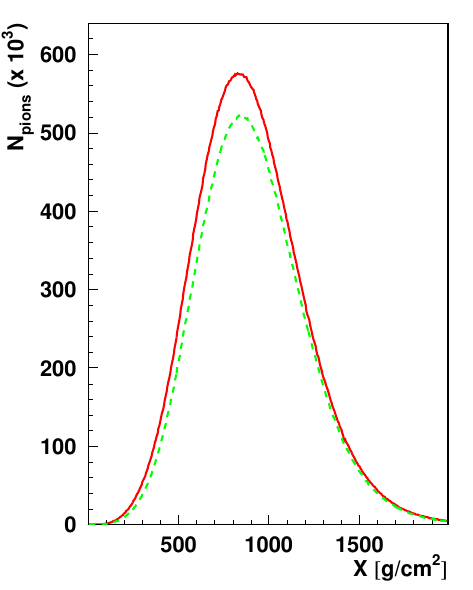}&
\includegraphics[width=2.8in]{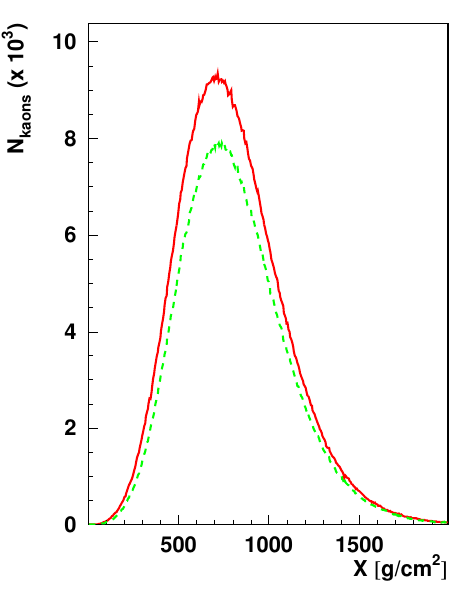}\\
\includegraphics[width=2.8in]{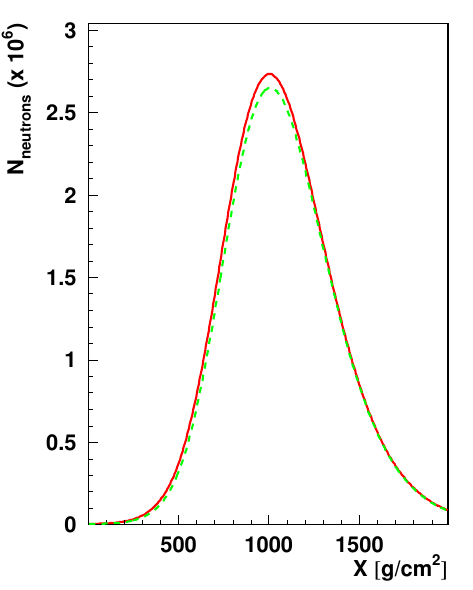}&
\includegraphics[width=2.8in]{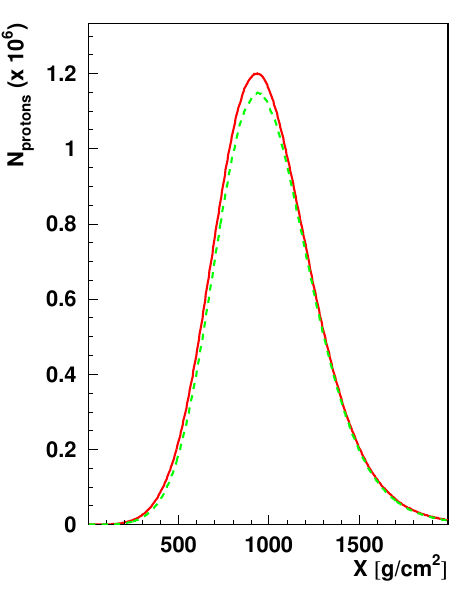}
\end{tabular}
\caption{Same as figure \ref{fig:longiepm} but for the longitudinal
  development of hadrons: Upper left pions, upper right kaons, lower
  left neutrons, lower right protons.}
\label{fig:longihad}
\end{figure}

The most probable photon interactions at the mentioned energy are
electromagnetic (i.e., pair production), and for that reason most of
the shower secondaries will be electrons and photons; and the
number of such secondaries is not expected to change substantially
when replacing the photonuclear cross sections. This can clearly be
seen in figure \ref{fig:longiepm} where the average shower
longitudinal development of electrons and positrons 
plotted show almost no differences between
the old and present models.
\begin{figure}
\begin{center}
\begin{tabular}{cc}
\includegraphics[width=2.8in]{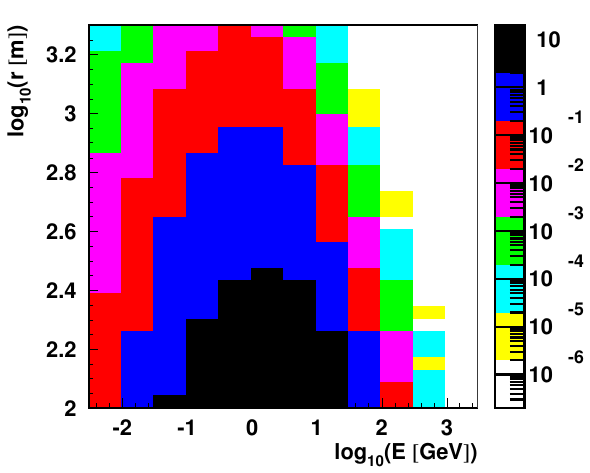} &
\includegraphics[width=2.8in]{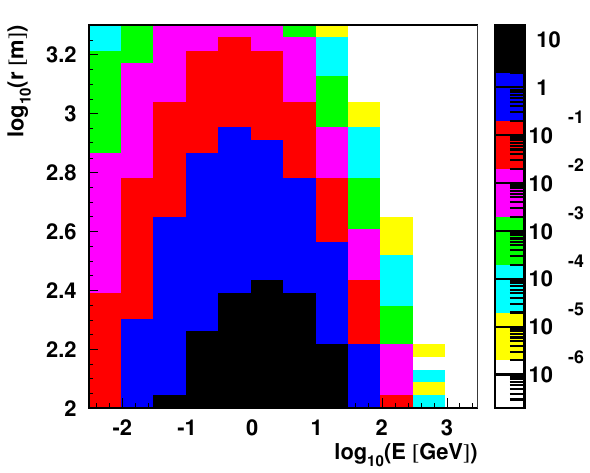}
\\
\includegraphics[width=2.8in]{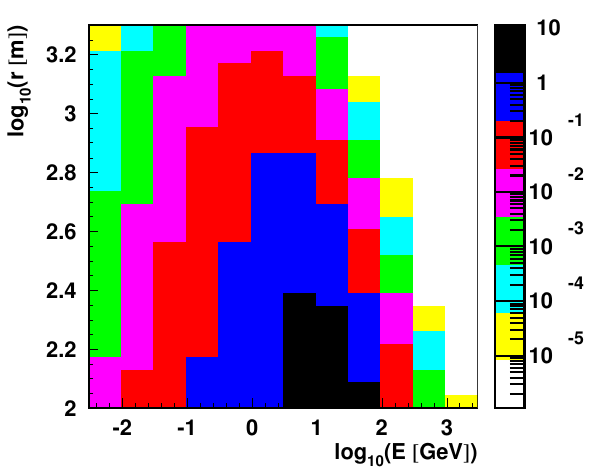} &
\includegraphics[width=2.8in]{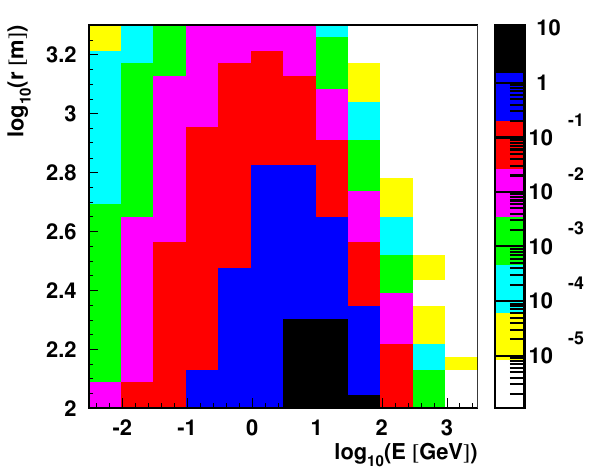}
\\
\includegraphics[width=2.8in]{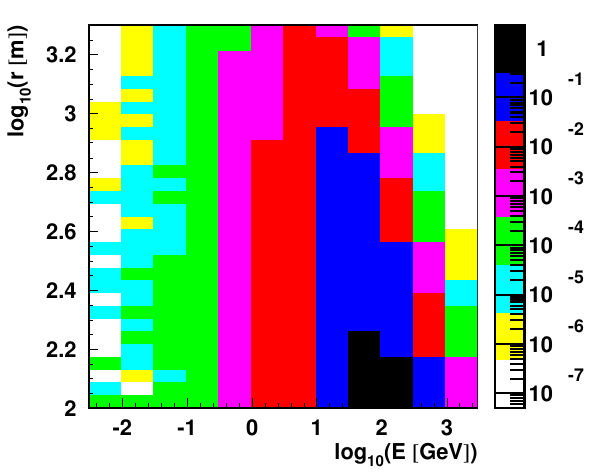} &
\includegraphics[width=2.8in]{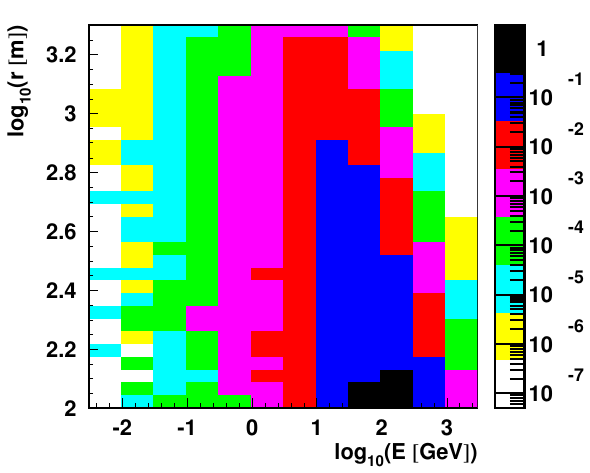}
\end{tabular}
\end{center}
\caption{Energy versus lateral (3d) distance to the shower axis
  distribution of ground muons for $10^{19}$ eV photon showers. The
  normalized density of muons $d\rho_\mu/d \log_{10}E$ (in m$^{-2}$)
  is represented in a color scale. The left (right) column plots
  correspond to simulations with the present (old) model for
  photonuclear cross sections. The upper, middle, and lower row plots
  correspond to zenith angles of 45, 60, and 80 deg respectively.}
\label{fig:r0egymudists}
\end{figure}

On the other hand, muon production is noticeably increased when
using the new photon cross sections. We present our results for the
longitudinal development of muons in figure \ref{fig:longimu}, where
it is clearly seen that the simulations with the present model produce more
muons in virtually the entire shower life. The relative difference
with respect to the old model is about 12$\,$\% at the maximum ($X\simeq
1100$ g/cm$^2$). The difference persists even in the very late stage
of shower development, as it can clearly be seen in the case of 80 degrees
inclined showers displayed at the right side of figure
\ref{fig:longimu}.
\begin{figure}
\begin{center}
\begin{tabular}{cc}
\includegraphics[width=2.8in]{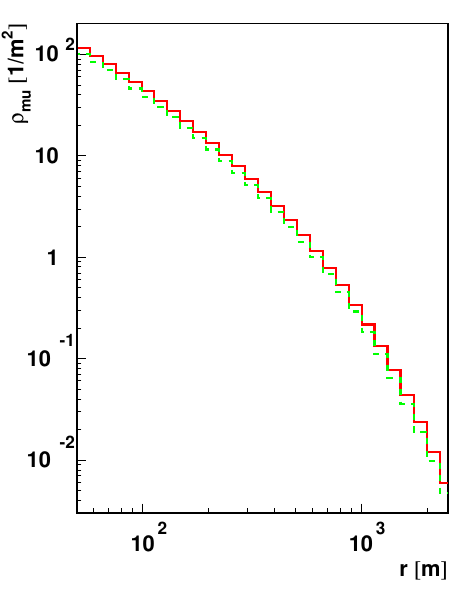} &
\includegraphics[width=2.8in]{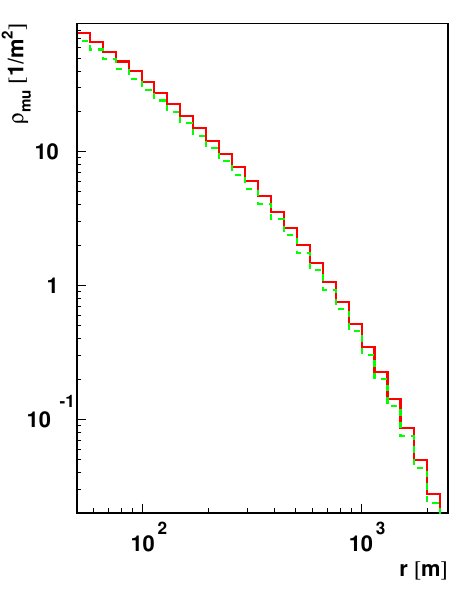}
\\
\includegraphics[width=2.8in]{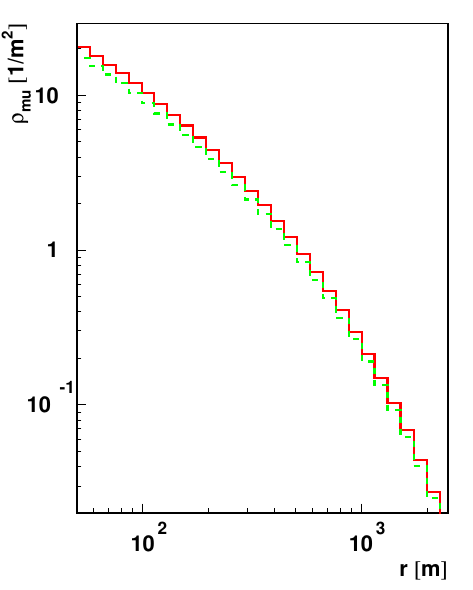} &
\includegraphics[width=2.8in]{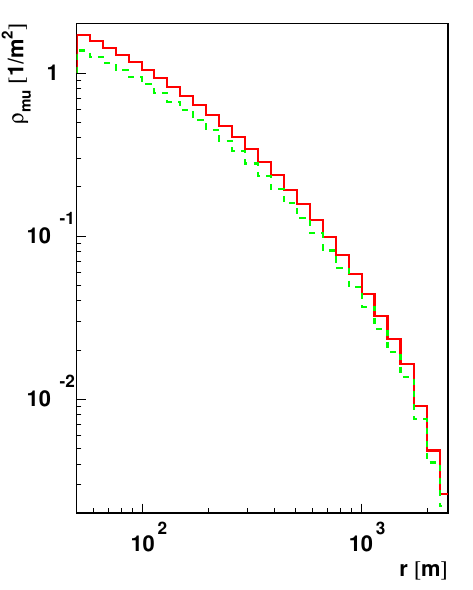}
\end{tabular}
\end{center}
\caption{Lateral distributions of ground
  muons for $10^{19}$ eV photon showers inclined 0 (upper-left), 45
  (upper-right), 60 (lower-left), and 80 (lower-right) degrees. The solid
  (dashed) line corresponds to simulations with the present (old)
  model for photonuclear cross sections.}
\label{fig:latmudistvangles}
\end{figure}
\begin{figure}
\begin{center}
\begin{tabular}{cc}
\includegraphics[width=2.8in]{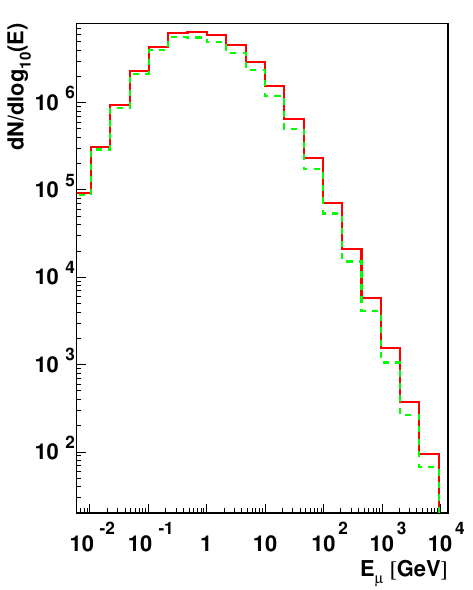} &
\includegraphics[width=2.8in]{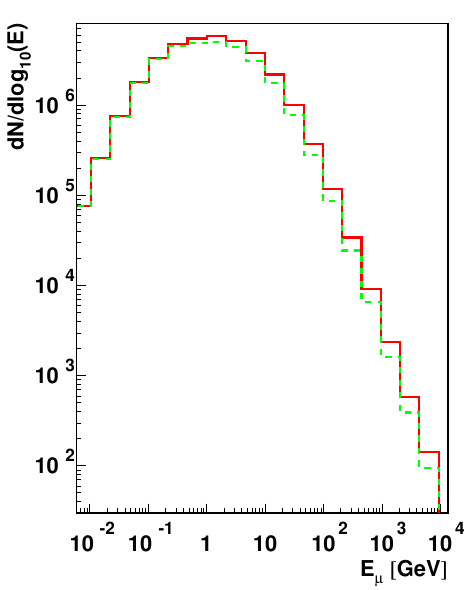}
\\
\includegraphics[width=2.8in]{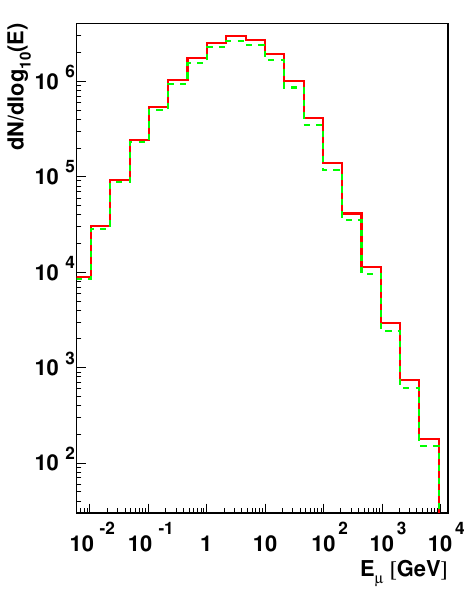} &
\includegraphics[width=2.8in]{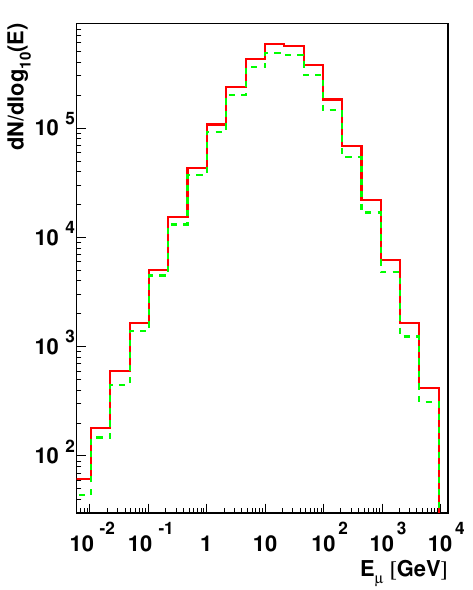}
\end{tabular}
\end{center}
\caption{Energy distributions of ground
  muons for $10^{19}$ eV photon showers inclined 0 (upper-left), 45
  (upper-right), 60 (lower-left), and 80 (lower-right) degrees. The solid
  (dashed) line corresponds to simulations with the present (old)
  model for photonuclear cross sections.}
\label{fig:egymudistvangles}
\end{figure}

It is important to recall that shower muons are generated after the
decay of unstable hadrons, mainly charged pions and kaons. Hence, the
enlarged number of muons that shows up in figure \ref{fig:longimu}
would be necessarily connected with enlarged hadron production,
especially pions and kaons. The results of our simulations agree with
this expectation. The results for the longitudinal development
of charged pions, charged kaons, neutrons, and protons, plotted in
figure \ref{fig:longihad} for the representative case of $10^{19}$ eV
gamma showers inclined 60 degrees illustrate this point. These plots
reveal a significant increase in the average number of produced pions
and kaons, when comparing the simulations performed with the present
model with the ones run with the old model. The plots for neutrons and
protons also indicate noticeable but smaller increments.  The current
figures are obtained using QGSJET-II to simulate hadronic
interactions. Simulations performed with other hadronic collision
packages could give numerically different results, but with the
characteristic that larger hadron production will always be expected
in the case of the present mode because of its increased gamma-nucleus
cross section, which enlarges the probability of hadronic collisions,
especially for very energetic photons, present mostly at the early
stages of shower
  development. The secondary particles generated after that initial
shower multiplication process are the ones recorded in the plots of
figure \ref{fig:longihad}, and their number will be increased every
time there is an increased hadronic collision probability.

It is also important to consider the characteristics of the muons
produced in the simulations, especially those that reach the ground
level. We will focus on the representative case
of $10^{19}$ eV gamma showers with ground altitude 1400
m.a.s.l. This corresponds roughly to an atmospheric slant depth of 900
g/cm$^2$. Accordingly with the results displayed in figure \ref{fig:longimu},
this depth is located short before the maximum of the muon
longitudinal profile.
\begin{figure}
\begin{center}
\includegraphics{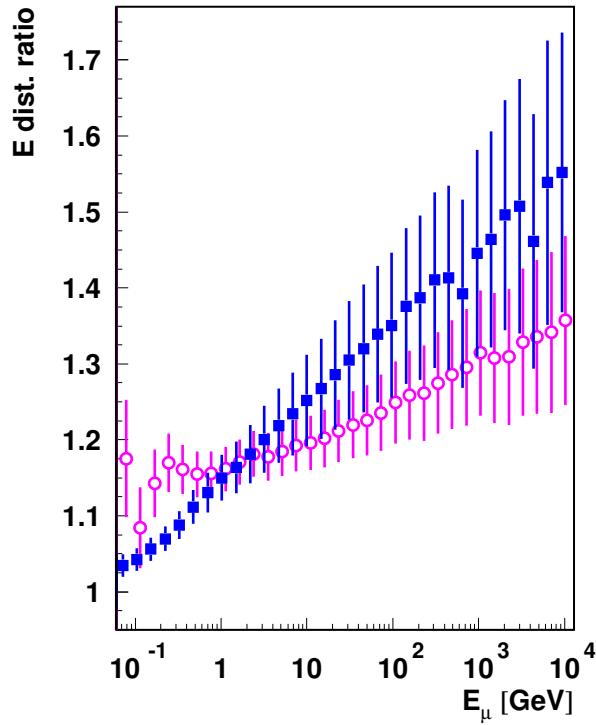}
\end{center}
\caption{Ratio between ground muon energy distributions obtained with
  the present and old models, for $10^{19}$ eV photon showers. The
  solid squares (open circles) correspond to a shower inclination of 45
  (80) deg. Error bars are calculated by propagation of the individual RMS
statistical errors of each of the distributions. The abscissas of the 80 
deg data set have been shifted by
  10\% to improve error bar visibility.}
\label{fig:muegydistratio4580}
\end{figure}

In figure \ref{fig:r0egymudists} the two dimensional, energy versus
3d distance to the shower axis, normalized density distribution of ground
muons is represented in a variety of cases. The left (right) column
show the distributions obtained from simulations performed using the
present (old) model for photonuclear cross sections. The upper,
middle, and lower row plots correspond to shower inclinations with
respect to the vertical of 45, 60, and 80 degrees, respectively. The
color scales used to represent the muon densities are unique at each
row. Comparing the distributions corresponding to the different
inclinations it shows up clearly that the number of ground muons
diminishes and their energy spectrum hardens as long as the zenith
angle is increased (notice the different color scales used at each
angle). Needless to say, this is the expected behavior for showers of
varying inclination, which at the same time will experiment a very
significant change in the ratio between the electromagnetic and muon
ground particle distribution (see for example reference
\cite{CillisSciutto2000}). When comparing the results corresponding to
simulations performed using the present and old models of photonuclear
cross sections (left and right column plots of figure
\ref{fig:r0egymudists}, respectively), it is possible to notice that
the densities of muons corresponding to the present model are larger
than the respective ones for the old model. The differences are
approximately independent of muon energy and distance to the shower
axis, as will be discussed in more detail in the following paragraphs.

The lateral and energy distribution of ground muons for various shower
inclination angles are displayed in figures \ref{fig:latmudistvangles}
and \ref{fig:egymudistvangles}, respectively. As in the case of the
distributions in figure \ref{fig:r0egymudists}, the simulations
correspond to $10^{19}$ eV gamma showers, with ground altitude of
1400 m.a.s.l.

In the case of the lateral distribution of muons (figure
\ref{fig:latmudistvangles}), we observe that the distributions for the
old and new photonuclear models are very similar in shape, differing
only in the total number of particles. It can also be observed that
the difference between photonuclear models becomes more significant
for large zenith angles.

On the other hand, the muon energy distributions displayed in figure
\ref{fig:egymudistvangles} present noticeable differences for muon
energies greater than roughly 1 GeV, with the present model giving the
largests number of particles at each bin. For muon energies lower than
1 GeV and zenith angles up to 60 degrees both distributions are
virtually coincident; in the case of showers inclined 80 degrees the
present model gives a larger number of particles in the entire muon
spectrum. To better illustrate this characteristic of the impact of
the photonuclear cross section on the average number of muons at
ground, we also include plots of the ratio between both muon energy
distributions. In figure \ref{fig:muegydistratio4580}, such ratios are
plotted as functions of the muon energy for the representative cases
of 45 (solid squares) and 80 (open circles) degrees of
inclination. The increased number of high energy muons resulting after
the simulations using the present model for photonuclear cross
sections shows up clearly in the case of showers inclined 45 degrees,
reaching average values of more than 50$\,$\% for muon energies of
$10^4$ GeV. In the case of showers inclined 80 degrees, the relative
difference is always below 35$\,$\%, and remains virtually constant at
15$\,$\% approximately for muon energies below 1 GeV.

\section{Final Remarks}

The main objective of this paper is to present a QCD-based
model for photoproduction, updated from the previous analysis
\cite{ourphoton} in light of recent LHC results for total $pp$
cross sections, and to study the impact of this new model on the
photon initiated air
shower development. This model produces a photon-air nucleus total
cross section significantly larger than the previous 
 model included
in the standard extended air shower studies. The present analysis
based on simulations using the AIRES system clearly shows that
for photon initiated showers the total muon production is increased in
a measurable way. This result could be of direct importance in future
determinations of bounds for the highest energy cosmic photon flux,
particularly in the case of very inclined showers whose analysis is
strongly based on ground muon distributions \cite{AugerMuons2015}.  In
this respect, a more detailed analysis of this kind of effects is in
progress.

\section*{Acknowledgments}

This collaboration was partially financed by the Programa de
Cooperaci\'on Cient{\'\i}fico Tecnol\'ogico Argentino-Espa\~nol:
MinCyT-MINCINN AIC10-D-607.  F.C. and A.G. 
acknowledge financial support from
Junta de Andaluc{\'\i}a (FQM-330, FQM-101, FQM-437, FQM6552) and MINECO
FPA2013-47836-C3-1-P and Consolider-Ingenio 2010 program CPAN
(CSD2007-00042). Partial support by CONICET and ANPCyT, Argentina, is
also acknowledged.


%
%

\begin{thebibliography}{99}
%
\bibitem{PAO} See {\it www.auger.org.}
%
\bibitem{Aab:2014aea}
  The Pierre Auger Collaboration,
  \journal{Phys. Rev.}{D90}{122006}{2014}.
%
\bibitem{Abraham:2009qb}
  The Pierre Auger Collaboration,
  \journal{Astroparticle Physics}{31}{399–406}{2009}.
%
\bibitem{TOTEM7} G. Antchev, et al., 
  \journal{Europhys. Lett.}{101}{21004}{2013}.
%
\bibitem{TOTEM8} G. Antchev, et al.,
  \journal{Phys. Rev. Lett.}{111 (1)}{012001}{2013}. 
%
\bibitem{ourphoton} R. M. Godbole, A. Grau, G. Pancheri and Y. N.
  Srivastava, \journal{Eur. Phys. J.} {C63} {69-85}{2009}
%
\bibitem{AIRES} S. J. Sciutto,
  \journal{Proc. 27th ICRC (Hamburg)}{1}{237}{2001};\hfil\break
  see also {\it www2.fisica.unlp.edu.ar/aires.}
%
\bibitem{ourproton} A. Corsetti, A. Grau, G. Pancheri and Y. N.
  Srivastava, \journal{Phys. Lett}{B382}{282}{1996};
  A. Grau, G.Pancheri, Y. N. Srivastava,
  \journal{Phys. Rev.}{D60}{114020}{1999};
  R.~M.~Godbole, A.~Grau, G.~Pancheri and Y.~N.~Srivastava,
  \journal{Phys. Rev.}{D72}{076001}{2005}. 
%
\bibitem{BN}
  F. Bloch and  A. Nordsieck,  \journal{Phys.Rev.}{52}{54}{1937}.
%
\bibitem{cline} D. Cline, F. Halzen and J. Luthe,
  \journal{Phys. Rev. Lett.}{31}{491}{1973}.
%
\bibitem{ISR}
  R. Biancastelli, C. Bosio, G. Matthiae , J. V. Allaby, W. Bartel,
  G. Cocconi, A. N. Diddens, R. W. Dobinson, A. M. Wetherell,
  \journal{Phys. Lett.}{B44}{113}{1973}.
%
\bibitem{CRCold}
  G. B. Yodh, Y. Pal and  J. S. Trefil,
  \journal{Phys. Rev. Lett.}{28}{1005}{1972}.
%
\bibitem{Gaisser:1988ra}
T. K. Gaisser and T. Stanev, \journal{Phys. Lett.}{B219}{375}{1989}.
\bibitem{Durand:1998ax} L. Durand and H. Pi, \journal{Phys. Rev.} {D40}{1436}{1989}.

%
\bibitem{froissart} M.~Froissart, \journal{Phys. Rev.}{123}{1053-1057}{1961}.

\bibitem{GRV} M.~Gluck, E.~Reya, and A.~Vogt,
  \journal{Z. Phys.}{C53}{127--134}{1992};
  \journal{Z. Phys.}{C67}{443--448}{1995};
  \journal{Eur. Phys. J.}{C5}{461--470}{1998}.
%
\bibitem{MRST} A.~D. Martin, R.~G. Roberts, W.~J. Stirling, and
  R.~S. Thorne, \journal{Phys. Lett.}{B531}{216--224}{2002}.
%
\bibitem{CTEQ} H.L. Lai, J. Botts, J.  Huston, J. G. Morfin,
  J. F. Owens, Jian-wei Qiu, W. K. Tung, H. Weerts,
  \journal{Phys. Rev.}{D51}{4763--4782}{1995}.
%
\bibitem{GRVPHO} M.~Gl\"uck, E.~Reya and A.~Vogt,
  \journal{Phys. Rev.}{D46}{1973}{1992}.
%
\bibitem{GRS} M. Gl\"uck, E. Reya and I. Schienbein,
  \journal{Phys. Rev.}{D60}{054019}{1999};
  Erratum, \journal{ibid}{D62}{019902}{2000}.
%
\bibitem{CJKL} F. Cornet, P. Jankowski, M. Krawczyk and A. Lorca,
  \journal{Phys. Rev.}{D68}{014010}{2003}.

%
\bibitem{corsetti1996} A. Corsetti, A. Grau, G. Pancheri and Y.N. Srivastava,
 \journal{Phys. Lett.}{B382} { 282-288}{1996}.

%
\bibitem{greco} M. Greco and P.Chiappetta,
  \journal{Nucl. Phys.}{B221}{269}{1983}.

%
\bibitem{ourlastPRD} D.A. Fagundes, A. Grau, G. Pancheri, Y.N. Srivastava and O. Shekhovtsova,
 \journal{Phys. Rev.}{D91}{114011}{2015}.
 %
%
%
\bibitem{ourfroissart} 	
A.~Grau, R. M.~ Godbole , G.~Pancheri, Y. N.~ Srivastava, \journal{Phys. Lett.}{ B682}{55-60}{2009}.
%
\bibitem{ourlastPLB} A.~Achilli, R.~M.~Godbole, A.~Grau, R.~Hegde,
  G.~Pancheri and Y.~Srivastava,
  \journal{Phys. Lett. B}{659}{137}{2008}. 
%
\bibitem{AUGER2012} 
  The Pierre Auger Collaboration,
  \journal{Phys. Rev. Lett.}{109}{062002}{2012}. 
%
\bibitem{Fletcher} R. S.~Fletcher, T. K. Gaisser and F.~Halzen,
  \journal{Phys. Lett.}{B298}{442}{1993};
  \journal{Phys. Rev.}{D45}{377--381}{1992},
   Erratum, \journal{ibid}{D45}{3279}{1992}.


\bibitem{collins} J. C.~Collins and G. A.~ Ladinsky,
  \journal{Phys. Rev.}{D43}{2847}{1991}.
%


\bibitem{ATLAS} G. Aad {\it et al.} (ATLAS Collaboration), 
   \journal{Nucl. Phys.}{B889}{486}{2014}.

%
\bibitem{block2004}
M. M. Block and F. Halzen, 
\journal{Phys.Rev.} {D70} {091901}{2004}.
%
\bibitem{block2014} 	
  M. M. Block, L. Durand and P. Ha, 
 \journal{Phys. Rev.}{D89}{094027}{2014}. 

\bibitem{MSTW2008} A. Martin, W. Stirling, R. Thorne and G. Watt,
 \journal{Eur. Phys. J.}{C 63}{189}{2009}.
%

%
\bibitem{QGSJET} S. Ostapchenko,
 \journal{Nucl. Phys. Proc. Suppl.}{B151}{143}{2006}. 
%
\bibitem{PDG} K.A. Olive et al. (Particle Data Group),
\journal{Chin. Phys. C}{38}{090001}{2014}; see also {\it pdg.lbl.gov.} 
%
\bibitem{Billoir2001} P. Billoir {\it et al.,\/} for The Pierre Auger
  Collaboration, \journal{Proc. 27th ICRC (Hamburg)}{1}{718}{2001}.
%
\bibitem{CillisSciutto2000} A. Cillis, S. J. Sciutto,
  \journal{J. Phys. G}{26}{309}{2000}.
%
\bibitem{AugerMuons2015} The Pierre Auger Collaboration,
  \journal{Phys. Rev.}{D91}{032003}{2015}. 
%
\end{thebibliography}
\end{document}